\documentstyle[amstex,epsfig,referee]{mn}
\newcommand{\apj}{Astrophys. J.}

\newif\ifAMStwofonts

\ifoldfss
  \ifCUPmtlplainloaded \else
    \NewTextAlphabet{textbfit} {cmbxti10} {}
    \NewTextAlphabet{textbfss} {cmssbx10} {}
    \NewMathAlphabet{mathbfit} {cmbxti10} {} 
    \NewMathAlphabet{mathbfss} {cmssbx10} {} 
  \fi
  \ifAMStwofonts
    \ifCUPmtlplainloaded \else
      \NewSymbolFont{upmath} {eurm10}
      \NewSymbolFont{AMSa} {msam10}
      \NewMathSymbol{\upi}     {0}{upmath}{19}
      \NewMathSymbol{\umu}     {0}{upmath}{16}
      \NewMathSymbol{\upartial}{0}{upmath}{40}
      \NewMathSymbol{\leqslant}{3}{AMSa}{36}
      \NewMathSymbol{\geqslant}{3}{AMSa}{3E}

    \fi
  \fi
\fi 

\ifnfssone
  \newmathalphabet{\mathit}
  \addtoversion{normal}{\mathit}{cmr}{m}{it}
  \addtoversion{bold}{\mathit}{cmr}{bx}{it}
  \newmathalphabet{\mathbfit} 
  \addtoversion{normal}{\mathbfit}{cmr}{bx}{it}
  \addtoversion{bold}{\mathbfit}{cmr}{bx}{it}
  \newmathalphabet{\mathbfss} 
  \addtoversion{normal}{\mathbfss}{cmss}{bx}{n}
  \addtoversion{bold}{\mathbfss}{cmss}{bx}{n}
  \ifAMStwofonts
    \ifCUPmtlplainloaded \else
      \UseAMStwoboldmath
      \makeatletter
      \new@mathgroup\upmath@group
      \define@mathgroup\mv@normal\upmath@group{eur}{m}{n}
      \define@mathgroup\mv@bold\upmath@group{eur}{b}{n}
      \edef\UPM{\hexnumber\upmath@group}
      \new@mathgroup\amsa@group
      \define@mathgroup\mv@normal\amsa@group{msa}{m}{n}
      \define@mathgroup\mv@bold\amsa@group{msa}{m}{n}
      \edef\AMSa{\hexnumber\amsa@group}
      \makeatother
      \mathchardef\upi="0\UPM19
      \mathchardef\umu="0\UPM16
      \mathchardef\upartial="0\UPM40
      \mathchardef\leqslant="3\AMSa36
      \mathchardef\geqslant="3\AMSa3E
    \fi
  \fi
\fi 

\ifnfsstwo
  \DeclareMathAlphabet{\mathbfit}{OT1}{cmr}{bx}{it}
  \SetMathAlphabet\mathbfit{bold}{OT1}{cmr}{bx}{it}
  \DeclareMathAlphabet{\mathbfss}{OT1}{cmss}{bx}{n}
  \SetMathAlphabet\mathbfss{bold}{OT1}{cmss}{bx}{n}
  \ifAMStwofonts
    \ifCUPmtlplainloaded \else
      \DeclareSymbolFont{UPM}{U}{eur}{m}{n}
      \SetSymbolFont{UPM}{bold}{U}{eur}{b}{n}
      \DeclareSymbolFont{AMSa}{U}{msa}{m}{n}
      \DeclareMathSymbol{\upi}{0}{UPM}{"19}
      \DeclareMathSymbol{\umu}{0}{UPM}{"16}
      \DeclareMathSymbol{\upartial}{0}{UPM}{"40}
      \DeclareMathSymbol{\leqslant}{3}{AMSa}{"36}
      \DeclareMathSymbol{\geqslant}{3}{AMSa}{"3E}
    \fi
  \fi
\fi 

\ifCUPmtlplainloaded \else
  \ifAMStwofonts \else 
    \def\upi{\pi}
    \def\umu{\mu}
    \def\upartial{\partial}
  \fi
\fi

\title{Probing Solar Convection}
\author[]
       {M. Br\"uggen \\
        Max-Planck-Institut f\"ur Astrophysik,
Karl-Schwarzschild-Str.1, 85740 Garching, Germany\\ and \\Churchill
College, Cambridge, CB3 0DS, UK}

\pagerange{\pageref{firstpage}--\pageref{lastpage}}
\pubyear{1999}

\begin{document}

\maketitle

\label{firstpage}

\begin{abstract}
In the solar convection zone acoustic waves are scattered by turbulent
sound speed fluctuations. In this paper the scattering of waves by
convective cells is treated using Rytov's technique. Particular care is
taken to include diffraction effects which are important especially
for high-degree modes that are confined to the surface layers of the
Sun. The scattering leads to damping of the waves and causes a phase
shift. Damping manifests itself in the width of the spectral peak of
p-mode eigenfrequencies. The contribution of scattering to the line
widths is estimated and the sensitivity of the results on the assumed
spectrum of the turbulence is studied. Finally the theoretical
predictions are compared with recently measured line widths of
high-degree modes.
\end{abstract}

\begin{keywords}
convection, turbulence, waves, Sun: oscillations.
\end{keywords}

\section{INTRODUCTION}

In the solar convection zone thermal instabilities give rise to
turbulent convection as the predominant heat transport
mechanism. Turbulence is one of the unsolved puzzles of modern
physics, and the lack of a fundamental theory of turbulence requires a
variety of phenomenological models. This constitutes a serious
limitation to the predictiveness of many astrophysical models and very
often leads to a built-in uncertainty (Canuto \& Christensen-Dalsgaard
1996). Theoretical modelling of convection in the Sun is extremely
difficult. Solar convection is highly turbulent and occurs on a vast
range of scales. The material is compressible and magnetic fields can
play an important role. Consequently, various parametrizations have
been invented to describe the properties of the convection. At the
same time vast efforts are being undertaken to numerically simulate
solar convection. These simulations are still at an early stage and
their results are not yet conclusive. One of the prime motivations for
studying solar and stellar structure in great detail is to improve and
test these models of convection.\\

Statistically significant differences between the observed and
theoretically predicted eigenfrequencies of the Sun persist (Gough et
al. 1996). The differences between computed adiabatic eigenfrequencies
of the standard solar model and measured p-mode frequencies for any
given order $n$ increase with frequency when scaled with the mode
inertia (with the model frequencies being bigger).  Higher frequencies
of modes of a given radial order correspond to greater values of the
spherical degree $l$, and hence to modes that remain increasingly
confined to the surface layers of the Sun. This suggests that the
cause for the discrepancy between theory and observation lies in an
inadequate modelling of the surface layer of the Sun. It is this
surface layer where convection is important and constitutes the
greatest uncertainty in the model.\\

Helioseismic inferences are not reliable unless details of stellar
structure and wave propagation within the acoustic cavity are
correctly treated. It has been noted elsewhere (Canuto \&
Christensen-Dalsgaard 1996) that the errors in the computed
frequencies caused by uncertainties in the equation of state (EOS) are
large compared to the observational errors. The variation in the
average wave propagation properties inside the Sun caused by turbulent
convection and temperature fluctuations constitute yet another process
that will affect the eigenfrequencies. This influence is difficult to
assess and has not received as much attention as the errors introduced
by uncertainties in the EOS or the opacity. Since helioseismic
inferences are dependent on the detailed treatment of turbulence, the
Sun could serve as a laboratory to study turbulence under conditions
unattainable on Earth.\\

Turbulence enters the theory of solar oscillations in two places: it
changes the model of solar structure mainly through the turbulent
pressure that provides additional support against gravity, and
secondly it interacts with the waves themselves. The waves interact
with the turbulence via two mechanism: The turbulent velocity advects
the acoustic waves (Brown 1984; Swisdak \& Zweibel 1998) and the
temperature fluctuations cause fluctuations in the refractive index
and thus in the local phase speed of the waves. \\

The good quality of solar oscillation data does not only allow an
accurate determination of the eigenfrequencies but also contains 
detailed information such as the width of the spectral peak. 

The observed line width of acoustic modes is caused by absorption and
other non-adiabatic effects (e.g., Gough 1980; Goldreich \& Kumar
1991; Balmforth 1992), as well as by scattering by turbulent
velocity fluctuations in the solar convection zone.

The scattering contribution to the line width has been studied by
Goldreich \& Murray (1994) in the framework of normal modes. Using a
simple model of scattering of standing waves in a box and within the
strict limits of geometrical acoustics, they derived a scattering
width for radial modes of $\Gamma_{\rm s}\simeq \omega M^2/\pi(n+1)$,
where $\omega$ is the angular frequency of the mode, $M=u/c$ is the
Mach number of the turbulent velocity and $n$ the order of the mode.
However, geometrical acoustics is a poor approximation for non-radial
high-degree modes, because at the top of the acoustic cavity the
correlation length of the perturbations becomes comparable to the
wavelength of the waves.\\

In this paper the scattering contribution to the line width is
calculated, explicitly taking diffractive effects into
account. Recently measured line widths of high-degree modes are
compared with results of this calculation. Thus, one of the aims of
this study is to investigate to what extent the structure of solar
convection can be probed by seismic methods.  For this purpose a
formalism to describe the propagation of waves through random
fluctuating media is introduced. For the sake of clarity I
start by deriving the formalism for a homogeneous medium before
generalising it to inhomogeneous media in Sec.\,2.2.

\section{WAVE PROPAGATION THROUGH FLUCTUATING MEDIA}

Let us write down a wave equation with a scattering term on the rhs:

\begin{equation}
\hat{L}\psi=(\nabla^2+k^2)\psi=\epsilon V({\bf r})\psi,\label{con.2.1}
\end{equation}
where $V$ is sometimes called the `scattering potential'. For a
homogeneous medium $\epsilon V({\bf r})=2k^2\delta c({\bf r})/c({\bf
r})$, where $\delta c$ is the rms of the sound speed perturbation. As
the scattering is assumed to be weak, $\epsilon$ is a small
number. Thus one can solve equation (\ref{con.2.1}) perturbatively. An
approximate solution can be written in terms of a Born expansion
(e.g. Nayfeh, 1973) by writing

\begin{equation}
\psi=\sum_{m=0}^{\infty}\epsilon^m\psi_m .\label{con.2.2}
\end{equation}
Substituting into equation (\ref{con.2.1}) yields

\begin{equation}
(\nabla^2+k^2)(\psi_0+\epsilon\psi_1+...)=\epsilon V({\bf r})(\psi_0+\epsilon\psi_1+...).\label{con.2.3}
\end{equation}
Equating equal powers of $\epsilon$ yields for $m=0$:

\begin{equation}
\hat{L}\psi_0=0 \label{con.2.4}
\end{equation}
and recursively for $m>0$

\begin{equation}
\hat{L}\psi_m=V({\bf r})\psi_{m-1}.\label{con.2.5}
\end{equation}
Equations (\ref{con.2.4}) and (\ref{con.2.5}) can be solved
successively using Green's function

\begin{equation}
\psi_m({\bf r})=\int d^3{\bf r}_1 V({\bf r}_1)\psi_{m-1}({\bf r}_1)G({\bf r};{\bf r}_1) ,\label{con.2.6}
\end{equation}
where the Green function, $G$, is defined by

\begin{equation}
(\nabla^2+k^2)G({\bf r};{\bf r}_1)=\delta^3({\bf r}-{\bf r}_1).\label{con.2.7}
\end{equation}
For a homogeneous medium $G({\bf r};{\bf r}_1)=\exp (ik|{\bf
r}-{\bf r}_1|)/4\pi|{\bf r}-{\bf r}_1| $.  From equation
(\ref{con.2.6}) one can write

\begin{equation}
\psi_1({\bf r})= \int d^3{\bf r}_1 V({\bf r}_1) G({\bf r};{\bf r}_1)\psi_0({\bf r}_1),\label{con.2.8}
\end{equation}
and successively

\begin{equation}
\psi_m({\bf r})=\int d^3{\bf r}_1...\int d^3{\bf r}_mV({\bf r}_1)...V({\bf r}_m)G({\bf r}_2;{\bf r}_1)... G({\bf r};{\bf r}_m)\psi_0({\bf r}_1). \label{con.2.9}
\end{equation}

An alternative and possibly more intuitive technique to calculate
phase and amplitude fluctuations, known as Rytov's method, starts by
recasting the wavefield into an exponential (see, e.g., Ishimaru 1978):

\begin{equation}
\psi ({\bf r})={\rm e}^{\Psi({\bf r})}.
\end{equation}
Hence one can write
\begin{equation}
\nabla^2\psi =\psi [\boldsymbol{\nabla}\Psi\cdot\boldsymbol{\nabla}\Psi+\nabla^2\Psi],
\end{equation}
and equation (\ref{con.2.1}) becomes

\begin{equation}
\nabla^2\Psi+\boldsymbol{\nabla}\Psi\cdot\boldsymbol{\nabla}\Psi+k^2-\epsilon V =0 ,
\end{equation}
which is a nonlinear first-oder differential equation known as
Riccati's equation. Denoting $\Psi_0 :=\Psi$ in the absence
of fluctuations, i.e. when $V=0$, and writing $\Psi=\Psi_0 + \Psi_1$,
one obtains
\begin{equation}
\nabla^2\Psi_1+2\boldsymbol{\nabla}\Psi_0\cdot\boldsymbol{\nabla}\Psi_1 = - [\boldsymbol{\nabla}\Psi_1\cdot\boldsymbol{\nabla}\Psi_1-\epsilon V ].
\end{equation}
With the identity
\begin{equation}
\nabla^2(\psi_0\Psi_1)=(\nabla^2\psi_0)\Psi_1 + 2\psi_0\boldsymbol{\nabla}\Psi_1 + \psi_0\nabla^2\Psi_1 ,
\end{equation}
one obtains the following inhomogeneous equation for $\psi_0\Psi_1$

\begin{equation}
(\nabla^2 + k^2) (\psi_0\Psi_1) = [\boldsymbol{\nabla}\Psi_1\cdot\boldsymbol{\nabla}\Psi_1 - \epsilon V]\psi_0,
\end{equation}
which can be solved using Green's function, yielding

\begin{equation}
\Psi_1 ({\bf r})\simeq \frac{1}{\psi_0({\bf r})}\int G({\bf r}-{\bf r}')[\boldsymbol{\nabla}\Psi_1\cdot\boldsymbol{\nabla}\Psi_1 - \epsilon V]\psi_0({\bf r}')d^3{\bf r}' .
\end{equation}
To first order, $\Psi_1=0$ in the integrand, so that

\begin{equation}
\Psi_1 ({\bf r})\simeq -\frac{1}{\psi_0({\bf r})}\int \epsilon V G({\bf r}-{\bf r}')\psi_0({\bf r}')d^3{\bf r}' .
\end{equation}
Thus the first Rytov solution is given by

\begin{equation}
\psi ({\bf r}) = \psi_0 ({\bf r}){\rm e}^{\Psi_1({\bf r})} .\label{con.2.10a}
\end{equation}
It has been shown that the first term in the Rytov expansion is
superior to the first Born approximation (Keller 1969), and I shall
employ Rytov's technique in the following analysis. The two methods
discussed above, the Born series and Rytov's technique, are the most
commonly used techniques dealing with scalar wave propagation through
random media. More recently, Samelsohn \& Mazar (1996) have treated
this problem using a path-integral analysis on the basis of the
parabolic wave equation. For a detailed review of stochastic wave
propagation and scattering in random media, see, e.g. Ishimaru (1978)
and Klyatskin (1980).\\

Here it will be assumed that the sound speed fluctuations, and
consequently the $\Psi_1$ fluctuations are Gaussian with zero mean,
i.e. $\langle\Psi_1\rangle = 0$, where the brackets denote the mean
over time. The mean of equation (\ref{con.2.10a}) has the form of a
characteristic function and one can write
\begin{equation}
\langle {\rm e}^{\Psi_1}\rangle = {\rm
e}^{\textstyle{\frac{1}{2}}\langle\Psi_1^2\rangle}, \label{con.2.10b}
\end{equation}
where $\langle\Psi_1^2\rangle$ is the correlation function defined in
equation (\ref{con.3.3}) (see Munk \& Zachariasen 1976, Panchev 1971).\\

Whenever waves interact with obstacles or inhomogeneities in the
medium, diffraction effects may occur. Most problems in diffraction
theory cannot be solved exactly, so that a number of techniques have
been invented to find approximate solutions.  One of these techniques
is Huygen's principle, which states that every point of a wave front
can be considered as a source of secondary wavelets which mutually
interfere.  The application of this principle to diffraction problems
leads to the Fresnel zone construction.  The textbook example for the
application of Fresnel zones is the problem of diffraction by a small
aperture in an infinite screen as sketched in Fig.\,1. A wave
propagates from a point source at point S through an opening in an
opaque screen to point R.  According to Huygen's principle the total
disturbance at R due to a source at point S is given by the integral
over the aperture

\begin{equation}
\psi (R)=\int \frac{e^{ik(r+s)}}{r+s} dS,\label{Kirch}
\end{equation} 
where $s=\overline{SP}$ and $r=\overline{PR}$. In equation
(\ref{Kirch}) I have neglected a so-called obliquity factor, which
describes the angular variation of the secondary wavelets. Rays that
connect S and R via a point P in the plane now have a phase at R that
is different from the phase of the stationary ray.

The envelopes of all the rays of length $a+b+\lambda/2$,
$a+b+\lambda$, ..., $a+b+j\lambda /2$, divide the aperture into a
number of zones, which are called Fresnel zones ($a$ and $b$ are
defined in Fig.\,1, $\lambda$ is the wavelength and $j$ an
integer). The contributions from successive Fresnel zones to the total
wavefield alternate in sign. Those rays whose phase differs from the
stationary phase by $\pi$ form the boundary of the first Fresnel zone
in the plane of the obstacle. In our textbook example the first
Fresnel zone is a concentric circle centred on the optical axis. For a
more detailed treatment see, e.g., Born \& Wolf (1970).\\

Generally, the first Fresnel zone in a plane intersecting the raypath
is defined as the area bounded by all those rays whose phase differs
by $\pi$ from the phase of the stationary ray which joins two fixed
points. This can now be repeated for all planes intersecting the ray
path and the entirety of these zones yields a tube surrounding the
stationary ray path: this tube I call the Fresnel tube.  The Fresnel
zone provides an estimate of the size of an irregularity that would
give rise to diffraction effects. If the size of the irregularity is
much bigger than the Fresnel zone, diffraction effects are
negligible. It is straightforward to calculate the Fresnel zone for
waves in a homogeneous medium. Let me consider a wave propagating from
S to R as before. If the wave now follows the path SPR instead of the
stationary ray, the corresponding change in phase, $\phi$, is given by

\begin{equation}
\delta\phi = k[(a^2+\delta z^2)^{1/2}+[b^2+\delta z^2]^{1/2}-(a+b)]\simeq \frac{k(a+b)}{2ab}\delta z^2,
\end{equation}
where $k$ denotes $2\pi /\lambda$. Hence the size of the Fresnel zone
is given by

\begin{equation}
\sqrt{2\pi}\left (\frac{\partial^2\phi}{\partial z^2}\right
)^{-1/2}=\left [\frac{2\pi ab}{k(a+b)}\right ]^{1/2} .
\end{equation}

\subsection{Homogeneous medium}

I begin by considering the case of wave propagation in a fluctuating
homogeneous medium. In this case the ray paths are straight lines and
the turbulent correlation function is a function of separation only,
i.e.
\begin{equation}
\epsilon^2\langle V({\bf r}_1)V({\bf r}_2)\rangle =\zeta({\bf r}_1-{\bf r}_2).\label{con.3.1}
\end{equation}
As stated in equations (\ref{con.2.10a}) and (\ref{con.2.10b}) the
time-averaged wavefield at ${\bf r}$ due to a source at the origin is
given by the unperturbed wavefield times $\exp
({\textstyle\frac{1}{2}}\langle\Psi_1^2\rangle)$, where

\begin{multline}
\langle\Psi_1^2\rangle=(16\pi^2)^{-1}\int d^3{\bf r}_1\int d^3{\bf r}_2\frac{r^2}{r_1|{\bf r}-{\bf r}_1|r_2|{\bf r}-{\bf r}_2|}\zeta({\bf r}_1-{\bf r}_2)\\
\times \exp [ik(r_1+|{\bf r}-{\bf r}_1|-r)]\exp [ik(r_2+|{\bf r}-{\bf r}_2|-r)].\label{con.3.3}
\end{multline}
It is convenient to convert to relative and centre-of-mass coordinates:

\begin{equation}
{\bf \tilde{r}}={\bf r}_1-{\bf r}_2
\end{equation}
and

\begin{equation}
{\bf R}=\frac{1}{2}({\bf r}_1+{\bf r}_2).
\end{equation}
I assume that $\zeta$ is negligible when $\tilde{r}$ is bigger than the
correlation length of the sound speed perturbations, which are taken
to be smaller than $R$. Then one can expand the integrand in
(\ref{con.3.3}) in ${\bf \tilde{r}/R}$ yielding

\begin{multline}
\langle\Psi_1^2\rangle = \frac{1}{16\pi^2}\int\ d^3{\bf R}\frac{r^2}{R^2|{\bf r}-{\bf R}|^2}\exp [2ik(R+|{\bf r}-{\bf R}|- r)] \\
\int d^3{\bf \tilde{r}}\ \zeta ({\bf \tilde{r}})\exp\left [ \frac{ik}{4}\left ( \frac{{\bf \tilde{r}}^2-({\bf \tilde{r}}\cdot \hat{\bf R})^2}{R}+\frac{{\bf \tilde{r}}^2-[{\bf \tilde{r}}\cdot({\bf \hat{r}}- {\bf \hat{R}})]^2}{|{\bf r}-{\bf R}|}\right )\right ],\label{con.3.6}
\end{multline}
where the hat denotes a unit vector. Assuming that the radius of the
Fresnel zone is smaller than $s$, i.e. $(r-s)/krs\ll 1$, one can
expand equation (\ref{con.3.6}) in $R_{\perp}/R_{\parallel}$ about its
stationary path, which is the straight line from 0 to ${\bf r}$. Here
$s$ stands for $R_{\parallel}$. The subscripts $\perp$ and $\parallel$
denote the vector components perpendicular and parallel to ${\bf
r}$. Thus one can write

\begin{equation}
R+|{\bf r}-{\bf R}|-r\simeq \frac{{\bf R}_{\perp}^2r}{2s(r-s)}\ .\label{con.3.7}
\end{equation}
Moreover,

\begin{equation}
\frac{{\bf \tilde{r}}^2-({\bf \tilde{r}}\cdot \hat{{\bf R}})^2}{R}\simeq\frac{{\bf \tilde{r}}_{\perp}^2}{s}+ {\rm O}\left ( \frac{\tilde{r}^2_{\perp}R_{\perp}}{s^2}\right )\label{con.3.7a}
\end{equation}
and

\begin{equation}
\frac{{\bf \tilde{r}}^2-[{\bf \tilde{r}}\cdot({\bf \hat{r}}- {\bf \hat{R}})]^2}{|{\bf r}-{\bf R}|}\simeq \frac{{\bf \tilde{r}}_{\perp}^2}{r-s}+{\rm O}\left ( \frac{\tilde{r}^2_{\perp}R_{\perp}}{s^2}\right ).\label{con.3.7b}
\end{equation}
Substituting (\ref{con.3.7}) into (\ref{con.3.6}) and evaluating the
integral over ${\bf R}_{\perp}$ by stationary phase, one finds that

\begin{equation}
\int\ d^3{\bf R}\frac{r^2}{R^2|{\bf r}-{\bf R}|^2}\exp [2ik(R+|{\bf r}-{\bf R}|- r)]\simeq \frac{i\pi}{k}\int_0^r \frac{r}{s(r-s)}\ ds .\label{con.3.8}
\end{equation}
Neglecting terms of order $\tilde{r}^2R_{\perp}/s^2$ in equation
(\ref{con.3.6}) one is left with the integral over ${\bf \tilde{r}}$
which after substitution of (\ref{con.3.7a}) and (\ref{con.3.7b})
becomes

\begin{equation}
\langle\Psi_1^2\rangle \simeq \frac{i}{16\pi k}\int_0^r ds\int d^3{\bf \tilde{r}}\ \zeta ({\bf \tilde{r}})\frac{r}{s(r-s)}\times\exp\left [\frac{ik}{4}\left (\frac{1}{s}+\frac{1}{r-s}\right ){\bf \tilde{r}}_{\perp}^2\right ] .\label{con.3.9}
\end{equation}
It is convenient to use the Fourier transform of the correlation
function, which is defined as 

\begin{equation}
\tilde{\zeta}({\bf q})=\int \zeta ({\bf \tilde{r}})e^{i{\bf q}\cdot{\bf \tilde{r}}}\ d^3{\bf \tilde{r}}.\label{con.3.10}
\end{equation}
Substituting the inverse tranform into equation (\ref{con.3.9}), using
the relation

\begin{equation}
\int \zeta (\tilde{r}_{\parallel},0) d\tilde{r}_{\parallel}=(2\pi)^{-2}\int \tilde{\zeta}(0,{\bf q}_{\perp})d^2{\bf q}_{\perp},
\end{equation}
and integrating over $d^2{\bf r}_{\perp}$ by stationary phase, one obtains:

\begin{equation}
\langle\Psi^2_1\rangle \simeq-\frac{1}{(4\pi)^2k^2}\int d^2{\bf q}_{\perp}\ \tilde{\zeta}(0,{\bf q}_{\perp})\int_0^r ds\ \exp\left [\frac{iq_{\perp}^2(s-r)s}{kr}\right ] . \label{con.3.11}
\end{equation}
The reader will recognise that the argument in the exponential in
equation (\ref{con.3.11}) is essentially the ratio of the size of
the Fresnel zone to the correlation length of the fluctuations. The
exponential takes account of the diffraction effects; its argument
vanishes if the size of the inhomogeneities is much bigger than the
Fresnel zone.  The geometrical acoustics result is recovered by
expanding the exponential in the previous equation. This is valid only
if

\begin{equation}
q_{\perp}^2(s-r)s/kr\ll 1,\label{con.3.12}
\end{equation}
which is the well-known Fresnel condition. Then the integral in
(\ref{con.3.11}) is simply

\begin{equation}
\langle\Psi_1^2\rangle \simeq -(2k)^{-2}r\int_0^r \zeta (r')\ dr',\label{con.3.13}
\end{equation}
which is the geometrical acoustics result, and, unlike
(\ref{con.3.11}), does not account for diffraction.

Here, for simplicity, I confine myself to isotropic turbulence so that the
three-dimensional Fourier transform becomes

\begin{equation}
\tilde{\zeta}(q)=\frac{4\pi}{q}\int^{\infty}_0 \zeta (r) r\sin qr\ dr.\label{con.3.14}
\end{equation}
This now completes the description of the propagation of waves through
a homogeneous turbulent medium, the main result being equation
(\ref{con.3.9}). This subsection serves to illustrate the evaluation
of equation (\ref{con.2.10a}), which is most clearly demonstrated in a
homogeneous medium. The Sun is, of course, inhomogeneous; hence in the
following subsection I will generalise this formalism to inhomogeneous
media following the same principles and methods that were outlined
above.

\subsection{Inhomogeneous medium}

For an inhomogeneous medium the correlation function $\zeta$ is a
function of position as well as relative separation, i.e. $\zeta=\zeta
({\bf\tilde{r}},{\bf R})$. Furthermore, in an inhomogeneous medium the
amplitude is no longer simply proportional to the inverse of the
distance from the source, as for spherical waves in a homogeneous
medium. In general, one can write the Green function as follows:

\begin{equation}
G({\bf r},{\bf r}_1)=A({\bf r},{\bf r}_1) \exp [iS({\bf r},{\bf r}_1)],\label{con.4.3}
\end{equation}
where the phase is given by

\begin{equation}
S({\bf r},{\bf r}_1)=\int_{\bf r}^{{\bf r}_1} {\bf k}[{\bf r}(s)]\cdot d{\bf s}\ .\label{con.4.4}
\end{equation}
It was shown by Munk \& Zachariasen (1976), for example, that the
`normalisation' can be expressed as

\begin{equation}
A({\bf r},{\bf R})=\frac{1}{4\pi}\left [ {\rm det}\frac{\partial}{\partial r_{\perp\,i}}\frac{\partial}{\partial R_{\perp\,j}}S({\bf r},{\bf R})\right ]^{1/2},\label{con.4.5}
\end{equation}
where the derivatives are evaluated at $r_{\perp}=R_{\perp}=0$. Now,
the exponent $\langle\Psi_1^2\rangle$ reads

\begin{multline}
\langle\Psi_1^2\rangle = \int d^3{\bf R}\left [\frac{A({\bf
r},{\bf R})A({\bf R},0)}{A({\bf r},0)}\right ]^2\int d^3{\bf
\tilde{r}}\ \zeta({\bf \tilde{r}},{\bf R})\exp [i(S({\bf r},{\bf R}+{\bf \tilde{r}/2})+S({\bf r},{\bf R}-{\bf \tilde{r}/2})\\+S({\bf R}+{\bf \tilde{r}/2},0)+S({\bf R}-{\bf \tilde{r}/2},0)-2S({\bf r},0))] .
\end{multline}
Expanding the exponent in powers of ${\bf \tilde{r}}$, the linear
terms vanish, so that one is left with the zero- and second-order term
in ${\bf \tilde{r}}$. Thus one can write

\begin{multline}
\langle\Psi_1^2\rangle \simeq \int d^3{\bf R}\left (\frac{A({\bf r},{\bf R})A({\bf R},0)}{A({\bf r},0)}\right )^2\exp [2i(S({\bf r},{\bf R})+S({\bf R},0)\\-S({\bf r},0))]\int d^3{\bf \tilde{r}}\ \zeta({\bf \tilde{r}},{\bf R})\exp [{\textstyle\frac{1}{4}}i\tilde{r}_i\tilde{r}_j C_{ij}({\bf R})],\label{con.4.6}
\end{multline}
where $C_{ij}$ denotes the `phase curvature' defined as

\begin{equation}
C_{ij}({\bf R})=\frac{\partial}{\partial R_i}\frac{\partial}{\partial R_j}[S({\bf r},{\bf R})+S({\bf R},0)]\ .\label{con.4.7}
\end{equation}
The argument of the exponential in (\ref{con.4.6}) is summed over the
indices $i$ and $j$, each of which can take the values 1, 2 and 3.
The phase curvature, $C_{ij}$ can be visualised as follows: Imagine
displacing a ray which leads from 0 to ${\bf R}$ and from ${\bf R}$
onwards to ${\bf r}$ by a small amount $\delta{\bf R}$ so that the
perturbed ray consists of a segment from 0 to ${\bf R}+\delta{\bf R}$
and a segment from ${\bf R}+\delta{\bf R}$ to ${\bf r}$. The second
derivative of the phase measured at ${\bf r}$ due to a source at the
origin with respect to perpendicular displacements at a point ${\bf
R}$ is called phase curvature. Since the ray follows a path of
stationary phase, one finds that $C_{ij}=0$ if $i\neq j$. Moreover, one
can choose coordinates relative to the stationary ray from 0 to ${\bf
r}$ such that the only non-zero components of $C$ are the second
derivatives in two perpendicular directions to the stationary ray at
any point. The phase curvature is closely related to the concept of
Fresnel zones and will be discussed further in Sec.\,\ref{sec:fresnel}.\\

Evaluating the integral over $d^3{\bf R}$ by the method of stationary
phase, selects the unperturbed path from 0 to ${\bf r}$ as in the
homogeneous case. The integral over $d^3{\bf \tilde{r}}$ can again be
simplified by introducing the Fourier transform of $\zeta$:

\begin{multline}
\langle\Psi_1^2\rangle \simeq -\int_0^{\bf r}ds\ \left [\frac{A({\bf
r},{\bf R}(s))A({\bf R}(s),0)}{A({\bf r},0)}\right ]^2 C_{11}^{-1/2}
C_{22}^{-1/2}\\
\int d^2{\bf q}_{\perp}(s)\tilde{\zeta}[{\bf q}_{\perp}(s),R(s)]\exp [i q_{\perp j}^2C_{jj}^{-1}({\bf R}(s))],\label{con.4.9}
\end{multline}
where the argument of the exponential is summed over $j$ ($j=1,2$). It
is found that it is an excellent approximation, instead of using the
normalisation of the Green function as given by equation
(\ref{con.4.5}), to simply write $A({\bf r},{\bf R})=(4\pi |{\bf
r}-{\bf R}|)^{-1}$, as in the homogeneous case. The deviations from
homogeneity become only important in the phase. It is verified
numerically (by using the correct Green function for a polytropic
envelope, see Fan et al. 1995) that equation (\ref{con.4.9}) can
approximately be written as

\begin{equation}
\langle\Psi_1^2\rangle \simeq -(4\pi)^{-2}\int_0^{\bf r}ds\,k(s)^{-2}\int d^2{\bf q}_{\perp}(s)\tilde{\zeta}[{\bf q}_{\perp}(s),R(s)]\times\exp [i q_{\perp j}^2C_{jj}^{-1}({\bf R}(s))],\label{con.4.10}
\end{equation}
which is the expression that one would obtain using the amplitudes of
the Green function for a homogeneous medium. Again equation
(\ref{con.4.10}) is much more accurate than the corresponding
expression in the geometrical limit, which is given by

\begin{equation}
\langle\Psi_1^2\rangle \simeq -\int_0^{\bf r}ds\ [2k(s)]^{-2}\int d{\tilde r}_{\parallel}\ \zeta ({\tilde r}_{\parallel},s) .
\end{equation}
Equation (\ref{con.4.9}) is the most important equation of this
paper. Within the assumptions made in its derivation, it describes the
interaction of waves with the turbulent medium. My aim is now to
evaluate (\ref{con.4.9}) for acoustic waves in the Sun. In the
following sections I will seek expressions for the various ingredients
of (\ref{con.4.9}), such as the scattering potential, the phase
curvature and the correlation of the turbulence.

\section{AN INHOMOGENEOUS WAVE EQUATION}

First I derive the wave equation for solar acoustic oscillations in the
presence of turbulence. This involves dividing the physical quantities
into statistical averages and random fluctuations. The averages
determine the mean properties of the envelope, whereas the
fluctuations describe the turbulent convection and form the
inhomogeneous term in the wave equation. For a sketch of the
derivation of the linearized adiabatic wave equation in the absence of
turbulence the reader is referred to the appendix.

Assuming that the Mach number of the turbulent flow is small, one can
linearize in the perturbed variables. For simplicity I neglect the
advection of the waves by the turbulent velocity and solely consider
the effect of the perturbations in the sound speed. The perturbed
quantities are denoted by a prime. As usual, $c$ denotes local sound
speed, $\rho$ matter density and $p$ pressure. Thus equation
(\ref{4.06}) becomes

\begin{equation}
(c+c')^2\frac{D\rho}{Dt}=\frac{Dp}{Dt},\label{con.5.1}
\end{equation}
with equations (\ref{4.01}) and (\ref{4.05}) unchanged. Now I repeat
the steps outlined in the appendix to eliminate $p'$ and $\rho '$ to
find an equation for the scalar $\chi :={\rm div}\bxi$, where $\bxi$
denotes the displacement of the fluid. Hence, in the presence of sound
speed perturbations to first order in $c'$, equation (\ref{4.17})
becomes

\begin{equation}
\frac{\partial^2{\bf\chi}}{\partial
t^2}=\nabla^2(c^2\chi-g_0\bxi\cdot{\bf n})-\nabla
(\Gamma\chi)\cdot{\bf n}+\nabla^2(2c\hspace{1mm}c'\chi)
-\nabla(H^{-1}2c\hspace{1mm}c'\chi)\cdot{\bf n}
\end{equation}
Again making the substitution $c^2\chi = \rho^{-1/2}\psi$,
and neglecting the buoyancy frequency, one obtains 

\begin{multline}
c^2\left ( \frac{\partial^2}{\partial t^2}+\omega_{\rm c}^2 \right )\psi -\nabla^2\psi=\nabla^2\psi\ 2\delta c +4({\bf n}\cdot\nabla \delta c)({\bf n}\cdot\nabla\psi) \\
+(2\nabla^2(\delta c)-2\delta c\ \omega_{\rm c}^2/c^2)\psi,
\end{multline}
where $\delta c$ is the fractional sound speed fluctuation $c'/c$. On
the lhs I have written the familiar wave equation, modified by the
cut-off frequency, $\omega_{\rm c}$, in the absence of sound speed
perturbations, and on the rhs I have written all the scattering terms
proportional to the perturbation $\delta c$. On the rhs I can now
substitute the unperturbed values for the spatial derivatives of
$\psi$ by treating the waves as locally plane,
i.e. $\nabla^2\psi=-k^2\psi=-\psi(\omega^2-\omega_{\rm c}^2)/c^2$ and
${\bf n}\cdot\nabla\psi=ik_z\psi$. This brings the wave equation into
the form of equation (\ref{con.2.1}) with a rhs of the form $\epsilon
V({\rm r})\psi$. The treatment of the waves as locally plane is based
on a wave-like decomposition of the normal modes. In this
approximation the modes are represented as standing waves which in
turn are formed by mutually interfering inward and outward propagating
waves. Strictly speaking, this approximation is only valid when the
order of the mode is large and many wavelengths fit in any scale
height of the background medium. In practice, however, it turns out
that this is a good approximation even for moderate order. For further
reference see the appendix of this paper and Gough (1993).\\

Before proceeding to apply this formalism to the propagation of
acoustic waves in the solar convection zone, I should recapitulate the
assumptions made in this section: Plane-parallel geometry was assumed;
as the rays that are most affected by convection are those confined to
the surface layers of the Sun, they will feel the curvature of the Sun
only as a small perturbation. Hence this assumption is fairly
accurate. As customary, the gravitational acceleration is treated as a
constant and the buoyancy frequency, $N$, is ignored. Furthermore,
perturbations in the gravitational acceleration ${\bf g}$ were
neglected ("Cowling"-approximation). For linearization to be valid,
the sound speed fluctuations and the Mach number of the turbulent flow
are assumed to be small. Here, I only considered the scattering by the
sound speed perturbations, i.e. by the perturbations in the refractive
index. For a complete treatment of the interaction of waves with the
turbulence one would also have to include those terms that arise from
the advection by the turbulent velocity.  Finally, for simplicity, I
ignore the spatial derivatives of the pressure scale height $H$ in the
expression of the cut-off frequency and simply write it as
$\omega_{\rm c}=c/2H$.

\section{FRESNEL TUBES}\label{sec:fresnel}

In this section the phase curvature, which was introduced in Sec.\,2,
is calculated. At each point P on the wave path one can calculate how
the phase of the wave due to a source at a fixed point Q varies when
the raypath is slightly displaced from its stationary path. The first
derivative of the phase with respect to small displacements of the ray
path is zero, as expected by Fermat's theorem. The second derivative,
also called `phase curvature', is defined in equation
(\ref{con.4.7}). It is inversely proportional to the square of the
linear extent of the Fresnel zone which was introduced in Sec.\,2.\\

Now, one can calculate the Fresnel tube for sound waves in the
Sun. For simplicity one may assume that the solar envelope can be
described by a plane-parallel polytrope. Then $c^2=c_0^2z$ is linear
with depth and $c_0$ is a constant. Indeed it is found that a
polytrope of index $\mu\simeq 3$ yields a good fit to the observed
frequencies. I should remark that this is not globally a good fit to
the envelope, which is closer to a polytrope with $\mu=3/2$, but
rather a consequence of the fact that the frequencies are dominated by
the uppermost layers of the Sun. The ray equations for a
plane-parallel polytrope can easily be solved and I briefly quote the
results here. If $x$ denotes the horizontal coordinate and $z$ depth
below the surface, a ray that originates at the origin ($x=z=0$) and
propagates in the positive $x$-direction satisfies

\begin{equation}
x=a\left [\sin^{-1}(z/a)^{1/2}-(z/a)^{1/2}(1-z/a)^{1/2}\right ],\label{con.6.1}
\end{equation}
where $a$ is the depth of the lower turning point,
$a=\omega^2/c_0^2k_x^2$, where $\omega$ is the angular frequency of
oscillation and $k_x$ is the horizontal wavenumber and a constant. The
phase is given by

\begin{equation}
\phi=2\omega a^{1/2}c_0^{-1}\sin^{-1}(z/a)^{1/2}.\label{con.6.2}
\end{equation}
So for the full arc between two photospheric reflections the ray has
traversed a horizontal distance $x=\pi a$ and acquired a phase
$\phi=2\omega(\pi x)^{1/2}/c_0$. In order to find the phase curvature
at any given point along the ray, one perturbs the ray at $x=x_0$,
while keeping the endpoints of the ray fixed. The second derivative of
the total phase with respect to this displacement of the ray is the
phase curvature as introduced in Sec.\,2. The vertical extent of the
first Fresnel zone is given by $\sqrt{2\pi}(\partial^2\phi /\partial
z^2)^{-1/2}$. In Fig.\,3 the extent of the first Fresnel zone in the
direction perpendicular to the ray is shown as a function of depth for
a ray with $a=0.001 R_{\odot}$. The Fresnel zone increases with depth
until approximately 2/3 of the depth of the lower turning point
from whereon it shrinks again to a smaller value at the lower turning
point (see also Jenson et al. 1998).\\

However, it should be noted that the lower turning point is a caustic
of the ray where the asymptotic ray theory breaks down. Nevertheless,
this does not considerably affect the calculation presented here,
since the damping of the waves occurs predominantly in the uppermost
layer of the solar envelope. In deeper regions, the turbulent Mach
number is too small to have any effect. If one assumes that the
convective cells have a typical size of a pressure scale height, one
can note that the size of the Fresnel zone is of the same order of
magnitude as the size of the convective cell. Therefore, diffractive
effects are important when the acoustic waves interact with the
convection.\\

Since the Sun is horizontally stratified (i.e. the sound speed being a
function of $z$ only), the phase curvature in the direction
perpendicular to the plane of the ray is the same as for a ray of the
same length in a homogeneous medium. The phase curvature for this case
was already calculated at the beginning of this section, and hence I
can write

\begin{equation}
\left .\frac{\partial^2\phi}{\partial y^2}\right |_{z=z_0}=\frac{2ka}{s(z_0)[2a-s(z_0)]},
\end{equation}
where $2a$ is the total length of a ray between two photospheric
reflections (in a plane-parallel polytrope), and $s(z_0)$ is the length
of the arc from 0 to ${\bf r}(z_0)$. 

\section{SIMPLE MODELS OF CONVECTION}

The simplest parametrization of convection is the mixing-length theory
(MLT) first devised by Taylor (1915, 1932). It describes turbulence by
a single length scale $l_{\rm m}$ which represents the dominant scale of
coherent motion. The mixing length $l_{\rm m}$ can be the characteristic
height of a convective cell or a mean free path. One imagines either
an ensemble of rising and falling turbulent elements that break up
after having traversed a distance $l_{\rm m}$ or a set of eddies of typical
diameter $l_{\rm m}$. Vitense (1953) adopted a mixing length that was proportional to
the pressure scale height, $l_{\rm m}=\alpha H$, where $H=-(\partial \ln
p/\partial r)^{-1}$ is the pressure scale height and $\alpha$ is the
proportionality constant which has to be determined by calibrating the
theory. This is done by evolving solar models with different values of
$\alpha$. The resulting one parameter family of mixing-length models
is then calibrated against the known radius of the Sun. Thus Gough \&
Weiss (1976) found a value of $\alpha =1.1$.\\

Ignoring pressure fluctuations the convective heat flux can be written as

\begin{equation}
F_{\rm c}\simeq\rho c_p \overline{wT'},\label{con.7.4}
\end{equation}
where the overbar denotes horizontal average. Now consider the
dynamics of the buoyant fluid. The vertical component of the turbulent
velocity $w$ can be estimated by equating the work done by the
buoyancy forces as the fluid rises through a height $l_{\rm m}$ to the kinetic
energy gained by the fluid, i.e.

\begin{equation}
\rho w^2 \simeq |\rho '|gl\simeq g\rho\delta l_{\rm m}|T'|/T,\label{con.7.5}
\end{equation}
where $\delta=-(\partial\ln\rho /\partial\ln T)_p$. Depending on the
geometry of the flow considered there may be another factor of order
unity in equation (\ref{con.7.5}).

When the convection is very efficient, thermal diffusion may be
ignored. The temperature fluctuations are then given by

\begin{equation}
|T'|\simeq\beta l_{\rm m},\label{con.7.6}
\end{equation}
where $\beta=-[dT/dz-(\partial T/\partial p)_{\rm
ad}dp/dz]$ is the superadiabatic gradient. Again I am neglecting
factors of order unity depending on the particular theory. Combining
equation (\ref{con.7.6}) with (\ref{con.7.5}) gives

\begin{equation}
w^2\simeq g\delta\beta l_{\rm m}^2/T.\label{con.7.7}
\end{equation}
To calculate the convective heat flux one notes that the temperature
fluctuation is positive for rising fluid and negative for falling
fluid, so that all motion contributes positively to $F_{\rm c}$. So
combining (\ref{con.7.6}) and (\ref{con.7.7}) with (\ref{con.7.4}) one
obtains

\begin{equation}
F_{\rm c}\simeq\rho c_p\overline{wT'}=Al_{\rm m}^2,\label{con.7.8}
\end{equation}
where $A=\rho c_p\beta^{3/2}(g\delta/T)^{1/2}$. This result was first
obtained by Prandtl (1932).

The sound speed perturbations are given by $|c'|/c=|T'|/2T$. $|T'|/T$
can be calculated using equation (\ref{con.7.5}), where the values of
$w$ and $l_{\rm m}$ are taken from a detailed model; the resulting
sound speed perturbations are shown in Fig.\,4.\\

Now one can extend the simple mixing-length approach beyond the
assumption that all eddies are of the same size, by introducing an
eddy spectrum or distribution, $\Phi(l_{\rm m})$, that describes the
density of eddies of size $l_{\rm m}$. In mixing-length theory
$\Phi(l_{\rm m})=\delta(l_{\rm m}-l_0)$, where $l_0$ now denotes the
mixing length of the standard MLT. Since there is no fundamental
theory of turbulence which would apply to conditions inside the Sun,
one has to choose a convenient spectrum. However, this eddy spectrum
has to be normalised to yield the same convective heat flux as
obtained by the calibrated mixing-length result. Therefore, $F_{\rm
c}$ is integrated over the entire eddy distribution to yield a total
convective heat flux of

\begin{equation}
F_{\rm c}=\int^\infty_0 Al_{\rm m}^2\Phi (l_{\rm m}) dl_{\rm m}.\label{con.7.11}
\end{equation}

For instance, one might decide to choose an eddy distribution of the form

\begin{equation}
\Phi (l_{\rm m})=B l_{\rm m} l_0^{-2}e^{-l_{\rm m}/l_0}.\label{con.7.12}
\end{equation}
Performing the integral in equation (\ref{con.7.11}) and equating it
to the calibrated MLT result, yields the normalisation to be $B=1/6$.

\section{RESULTS AND DISCUSSION}

Finally, I proceed to numerically evaluate the integrals in equation
(\ref{con.4.9}) using the phase curvature calculated in Sec.\,4 and
the turbulent spectrum given by equation (\ref{con.7.12}). The
integrals in the correlation function $\epsilon^2\langle V({\bf
r}_1)V({\bf r}_2)\rangle$ can be simplified by expanding the terms
that depend only on the static background state about their
centre-of-mass coordinate ${\bf R}$ assuming that all quantities
related to the static background do not vary substantially over a
correlation length of the turbulent perturbation.\\

The remaining correlation between the sound speed perturbations is
then taken to be

\begin{equation}
\langle\delta c({\bf r}_1)\delta c({\bf r}_2)\rangle \simeq \delta c^2({\bf R})\tilde{r}\Phi (\tilde{r}),
\end{equation}
and similarly for the correlations involving the derivative of $\delta
 c$. Here $\Phi (\tilde{r})$ is given by equation (\ref{con.7.12}) and
$\delta c({\bf R})$ is shown in Fig.\,4. In general, the
exponent $\langle\Psi_1^2\rangle$ is complex. The real part leads to
damping of the waves, and the imaginary part represents a phase
shift.\\

The real part of $\langle\Psi_1^2\rangle$ as a function of degree $l$
with the lower turning point kept fixed is shown in
Fig.\,5. Since the lower turning point is kept fixed,
frequency increases with increasing $l$. One can note that the real
part of $\langle\Psi_1^2\rangle$ is negative, which means that the
waves are damped, and the damping increases rapidly with $l$ (or
$\omega$). This does not come as a surprise as one would expect the
scattering to become stronger with decreasing wavelength. One may
quote, for example, scattering by small spheres (Rayleigh scattering),
where the scattering cross section is proportional to the fourth power
of $k$. Clearly, the scattering increases with the thickness of the
scattering medium and thus with increasing depth of the
lower turning point as demonstrated in Fig.\,5. Shown are
the results based on three different correlation functions for the
sound speed perturbations. The solid line assumes a spectrum of the
form given by equation (\ref{con.7.12}). The dashed line corresponds
to

\begin{equation}
\Phi (l_{\rm m})= (2 l_0)^{-1}e^{-l_{\rm m}/l_0},\label{con.7.14}
\end{equation}
and the dotted line is based on

\begin{equation}
\Phi (l_{\rm m})= l_{\rm m}^2(96 l_0^3)^{-1}e^{-l_{\rm m}/2l_0}.
\end{equation}
All three spectra are normalised according to equation
(\ref{con.7.11}). The dependence of the damping on the form of the
convective spectrum can be relatively significant. 

Fig.\,6 shows the imaginary part of $\langle\Psi_1^2\rangle$ for
lower turning points of depths $a=0.01\,R_{\odot}$ and
$a=0.001\,R_{\odot}$ as a function of degree. The results shown were
calculated on the basis of the spectrum given by equation
(\ref{con.7.12}). The phase shift is negative which implies that the
phase of the scattered wave is advanced over the phase of the
unperturbed wave. At first sight it may seem surprising that scattering
leads to an advancement of the phase. A simple qualitative explanation
for this result is given by Codona et al.  (1985). They showed that a
continuous random medium can cause an average advance of the arrival
time of a pulse, and I will briefly summarise their argument in the
following.\\

Consider a wave propagating through a homogeneous medium from a point
S to a point R, and assume that the random medium is concentrated in a
``phase screen'' at a distance $s$ from the source. This screen has
the effect of shifting the time of the wavefront by a small random
amount $t(x)$, where $x$ is the position on the screen and $t(x)$ is a
random distribution with zero mean. The travel time for a path through
$x$ is

\begin{equation}
\tau (x)=\tau_0+\frac{p}{2}\frac{x^2}{c}-t(x),
\end{equation}
where $\tau_0$ is the travel time of the unperturbed wave,
$p=r/s(r-s)$, $r$ being $\overline{\rm SR}$, and $c$ the wave speed in
the medium. By Fermat's principle the ray assumes a path such that
$\tau$ is stationary. Expanding $t(x)$ as follows

\begin{equation}
t(x)=t_0+t'x+{\textstyle\frac{1}{2}}t''x^2 + ... ,
\end{equation}
one finds that the ray crosses the screen at $x_{\rm s}=pct'$. The travel
time of the ray is then given by

\begin{equation}
\tau (x_{\rm s})\simeq \tau_0+{\textstyle\frac{1}{2}}cp{t'}^2-t_0-cp{t'}^2 \ .
\end{equation}
Since $t$, and hence $t_0$ and $t'$, are by construction random
variables with zero mean, the only contribution to the average travel
time comes from the squared terms. The first term is positive
corresponding to a delay, and represents the effect of geometry: The
perturbed path is geometrically longer than the unperturbed one. The
second term (called Fermat term) is negative and corresponds to a
pulse advance. It is also twice as big as the first one resulting in
an overall pulse advance. One can imagine that the rays governed by
Fermat's principle seek out regions with a pulse advance. Thus the
average travel time is given by
$\overline{\tau}=\tau_0-cp\overline{{t'}^2}/2$.  The phase shifts
displayed in Fig.\,6 are too small (corresponding to about 1/100th of
a second at a frequency of 3 mHz) to be detected directly as a time
delay in a time-distance analysis.\\

Having calculated the change of amplitude of the wave between two
photospheric reflections one can convert this attenuation into a
damping rate, $\eta$, using the equation $A/A_0=e^{-\eta\tau}$, where
$\tau$ is the travel time for a single traverse of the ray, and
$A/A_0$ the fractional change in amplitude per skip. Thus one finds
that $\eta=-\tau^{-1}\ln A/A_0$, where the amplitude ratio $A/A_0$ is
given by ${\rm
Re}[\exp(\textstyle{\frac{1}{2}}\langle\Psi_1^2\rangle)]$. Observationally
this damping manifests itself as a line width in the measured
frequency of a solar eigenmode. It is straightforward to show that the
line width (full width at half maximum) of the acoustic eigenmodes of
the Sun, $\Gamma$, is equal to twice the damping rate $\eta$.\\

With the advent of long time-series of solar oscillation data as
obtained, for example, by the ground-based GONG network and the
Michelson-Doppler Interferometer on board the spacecraft SOHO, the
line widths of high-degree modes can be measured with good accuracy.
Recently, line widths of high-degree modes were reported by Duvall,
Murawski, \& Kosovichev (1998) and their results for the p$_1$ modes
are reproduced in Fig.\,7 a. In Fig.\,7 b I have plotted the
fractional contribution to the line width as predicted by the method
presented here. The results shown in Fig.\,7 b are based on the
spectrum given by equation (\ref{con.7.14}). It is found that
scattering by sound speed perturbations contributes $\simeq$ 15\% to
the line width of the high-degree modes. However, the model presented
here can only be regarded as a toy model since the spectrum of the
turbulence and the model of the surface layers of the Sun were chosen
for convenience rather than for accuracy. But it is interesting to
note that this model correctly predicts the variation of the line
width with $\omega$ and $l$. The fractional contributions shown in
Fig.\,7 b are constant within the errors of the measurements.\\

Another caveat that I should mention concerns the assumed structure of
the turbulence. In Sec.\,5 the convection has been described by a
local mixing-length theory which presumes that the convection can be
characterised by a locally defined spectrum. This description of
turbulence was used in the computations in order to demonstrate the
capabilities of this method. However, recent numerical simulations and
laboratory experiments suggest that the local mixing-length picture is
a poor description of convection in stellar envelopes (e.g. Nordlund
\& Stein 1996). Instead there are indications that the turbulent flow
is extremely non-local, non-isotropic and hence poorly represented by
a local power spectrum.  Some simulations show that heat is
transported primarily through thin threads that extend all the way
from the bottom to the top of the solar convection zone. Nevertheless,
since Rytov's technique is an even better approximation if the
perturbations are confined to thin threads, the method of this
analysis will in principle remain valid and useful. \\

\section{CONCLUSIONS}

To summarise, I have presented an analytical description of the
interaction of waves with the turbulent medium following a similar
treatment used in oceanography. The inhomogeneous wave
equation was solved using Rytov's technique, and several of the integrals
in the ensuing expression could be evaluated analytically using a
stationary-phase approximation; the remaining integrals could be
solved numerically. Thus, in the framework of ray theory, one obtains
a quantitative assessment of the scattering contribution to the line
width of acoustic modes.\\

Previous studies of the line widths of solar modes have focused on
radial modes and generally neglected the effect of
diffraction. However, diffraction becomes important for shallow waves
for which the size of the Fresnel zone is comparable with the size of
the convective cells.

Scattering is not the only source of line widths. Other sources
include non-adiabatic effects associated with the radiative and
convective energy transport, mechanical absorption of the sound
waves (Gough 1980; Goldreich \& Keeley 1977; Christensen-Dals\-gaard,
Gough, \& Libbrecht 1989; Balmforth 1992) and partial reflection at
the upper turning point (Balmforth \& Gough 1990). The latter showed
in a simple analytical model that, for radial modes, partial
reflection contributes significantly to the line width. \\ 

In this paper I have isolated the effect of scattering by sound speed
perturbations from other effects that contribute to the line
width. Using a simple semi-analytical model I studied one of the
physical processes that determine the line widths of high-degree
modes. Once the remaining contributions to the line width are well
understood chances will be excellent that one will be able to
constrain the convective spectrum seismically.

\section*{Acknowledgments}
I am grateful to Douglas Gough, Henk Spruit and Takashi Sekii for
helpful discussions. The referee is thanked for many comments that 
improved the paper.

\newpage

\appendix

\section{Linearized adiabatic wave equation}

The linearized adiabatic wave equation has been derived by various
authors (e.g. see Gough 1986). In this appendix I briefly sketch its
derivation, which is referred to in Sec.\,3.\\

Neglecting viscosity, the momentum equation for a fluid moving with
velocity ${\bf u}$ can be written in the form

\begin{equation}
\rho\frac{D{\bf u}}{Dt}=-{\bf\nabla} p + {\bf g}\rho + \cal{F}, \label{4.01}
\end{equation}
where the Lagrangian derivative $D/Dt$ is defined as

\begin{equation}
\frac{D}{Dt}:=\frac{\partial}{\partial t}+{\bf u}\cdot{\bf\nabla}.\label{4.02}
\end{equation}
$\cal{F}$ denotes all other forces except gravity, $p$ is pressure and
$\rho$ matter density. ${\bf g}$ is the gravitational acceleration
which is related to the gravitational potential $\phi$ by

\begin{equation}
{\bf g}={\bf\nabla} \phi,\label{4.03}
\end{equation}
where $\phi$ satisfies Poisson's equation

\begin{equation}
\nabla^{2}\phi=-4\pi G\rho ,\label{4.04}
\end{equation}
with $G$ being the gravitational constant.\\

The continuity equation is given by

\begin{equation}
\frac{D\rho}{Dt}+\rho{\bf\nabla}\cdot{\bf u}=\frac{\partial\rho}{\partial t}+{\bf\nabla}\cdot(\rho{\bf u})= 0 . \label{4.05}
\end{equation}
Moreover, I make the assumption that the changes in pressure and
density are adiabatic, i.e.

\begin{equation}
\frac{Dp}{Dt}=\frac{\Gamma_1 p}{\rho}\frac{D\rho}{Dt}=c^{2}\frac{D\rho}{Dt}.\label{4.06}
\end{equation}
Here $\Gamma_1$ denotes the first adiabatic exponent, which is defined as 

\begin{equation}
\Gamma_1\equiv \left ( \frac{\partial \ln p}{\partial \ln \rho}\right )_{s}, \label{4.07}
\end{equation}
and which needs to be derived from an assumed equation of state; $c$
denotes the local sound speed.\\

Now, equations (\ref{4.01}), (\ref{4.04}), (\ref{4.05}) and
(\ref{4.06}) are perturbed around the equilibrium state in terms of
the displacement $\boldsymbol{\xi}$ of the fluid.  Perturbing the
equilibrium state -- $p=p_{0}+p'$, $\rho=\rho_{0}+\rho '$, and
$\phi=\phi_{0}+\phi '$ -- and by eliminating $p'$ and $\rho '$
from the momentum equation (\ref{4.01}) using equations (\ref{4.04})
and (\ref{4.05}), and after some rearrangement one obtains

\begin{equation}
\frac{\partial^{2}\boldsymbol{\xi}}{\partial t^{2}} = {\bf \nabla}(c^{2}\chi -g{\bf n}\cdot\boldsymbol{\xi})-\Gamma\chi {\bf 
n} \label{4.17}
\end{equation}
with $\Gamma=H^{-1}c^{2}-g$, where $H$ is the density scale height
$H:=-(\partial\ln\rho/\partial\ln r)^{-1}$. The scalar $\chi$ is defines as
$\chi:={\rm div}\boldsymbol{\xi}$.

Equation (\ref{4.17}) is an equation for the components of a
vector. However, a vector is coordinate dependent so that it would
be more convenient to find an equation for the scalar $\chi$.  This is
achieved by first taking the divergence of (\ref{4.17}):

\begin{equation}
\frac{\partial^{2}\chi}{\partial t^{2}} = \nabla^{2}(c^{2}\chi -g{\bf n}\cdot\boldsymbol{\xi})-{\bf n}\cdot{\bf \nabla}(\Gamma\chi) .\label{4.20}
\end{equation}
The vertical component of $\boldsymbol{\xi}$, ${\bf
n}\cdot\boldsymbol{\xi}$, can be reexpressed by taking the double curl
of equation (\ref{4.17})

\begin{equation}
{\bf n}\cdot\frac{\partial ^{2}}{\partial t^{2}}{\bf
\nabla}\times({\bf \nabla}\times\boldsymbol{\xi})=\frac{\partial ^{2}}{\partial
t^{2}}({\bf n}\cdot{\bf \nabla}\chi- \nabla^{2}{\bf n}\cdot\boldsymbol{\xi})=-g\nabla^{2}_{\rm h}(\Gamma\chi),\label{4.21}
\end{equation}
where $\nabla^{2}_{\rm h}$ is the horizontal Laplace operator. Solving
(\ref{4.21}) for ${\bf n}\cdot\boldsymbol{\xi}$ and substituting into (\ref{4.20})
yields the fourth-order equation:

\begin{equation}
\frac{\partial^{4}\chi}{\partial t^{4}} - \frac{\partial^{2}}{\partial
t^{2}}[\nabla^{2}(c^{2}\chi) - {\bf n}\cdot{\bf
\nabla}(H^{-1}c^{2}\chi)]-N^{2}
\nabla^{2}_{\rm h}(c^{2}\chi)=0 ,
\end{equation}
where $N$ is the buoyancy frequency given by 

\begin{equation}
N^2=g\left (\frac{1}{H}-\frac{g}{c^2}\right ) .
\end{equation}
Eliminating odd derivatives of the dependent variable by the
substitution $c^{2}\chi=\rho^{-1/2}\psi$, yields

\begin{equation}
c^{-2}\left ( \frac{\partial ^{2}}{\partial t^{2}}+\omega_{\rm c}^{2}\right ) \frac{\partial ^{2}\psi}{\partial t^{2}}-\frac{\partial ^{2}}{\partial t^{2}}\nabla^{2}\psi-N^{2}\nabla^{2}_{\rm h}\psi=0, \label{4.23}
\end{equation}
where the quantity $\omega_{\rm c}=\frac{c}{2H}(1-2{\bf n}\cdot{\bf
\nabla}H)^{1/2}$ is the critical cut-off frequency. For acoustical
oscillations one can simplify equation (\ref{4.23}) by neglecting the
buoyancy frequency $N$:

\begin{equation}
\left ( \frac{\partial ^{2}}{\partial t^{2}}+\omega_{\rm c}^{2}\right )\psi -c^{2}\nabla^{2}\psi=0. \label{4.24}
\end{equation}
Aside from the term containing the critical cut-off frequency this is
essentially a wave equation. The cut-off frequency $\omega_{\rm c}$
modifies the wave equation in the sense that in the region where
$\omega<\omega_{\rm c}$ the waves become evanescent.

\newpage

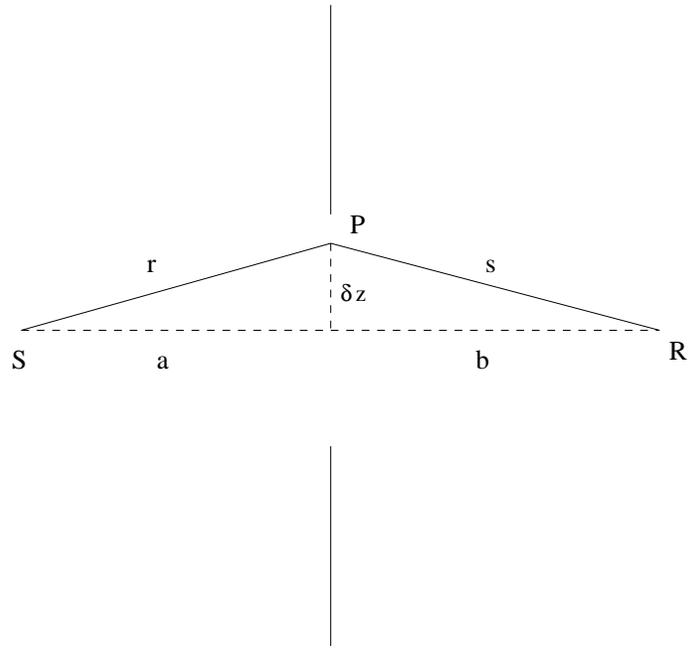
\begin{figure}
\caption{Fresnel zone construction.}
\label{fresneldiag}
\end{figure}

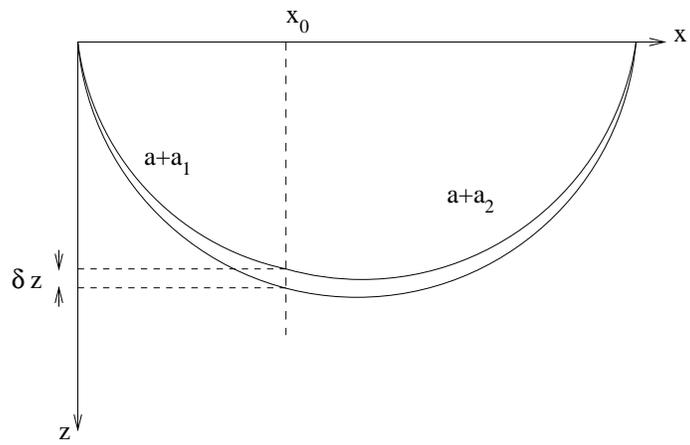
\begin{figure}
\caption{Diagram to the calculation of the Fresnel tube in the Sun.}
\label{fresneldiag}
\end{figure}

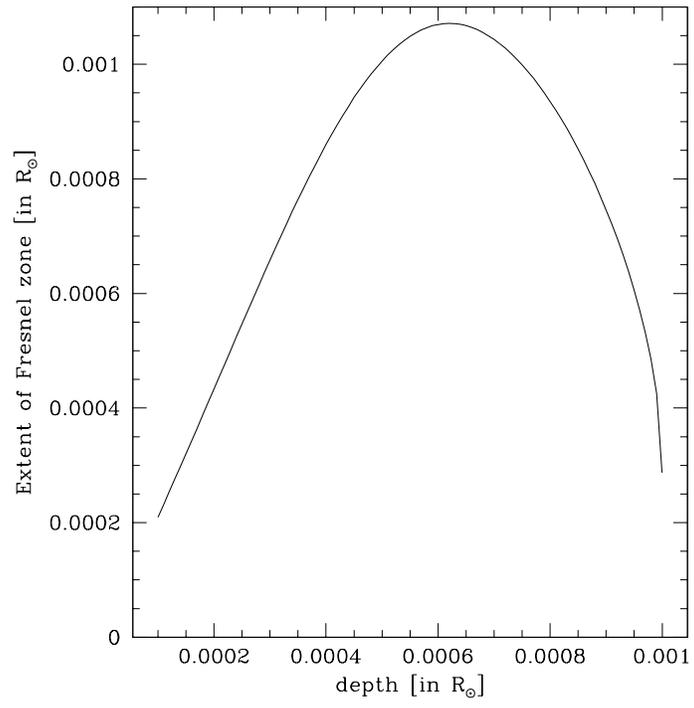
\begin{figure}
\caption{Extent of the Fresnel tube (perpendicular to
the ray) as a function of depth in a
plane-parallel polytrope for a ray with $a=0.001$. All lengths are
measured in units of solar radii.}
\label{Fresneltube}
\end{figure}

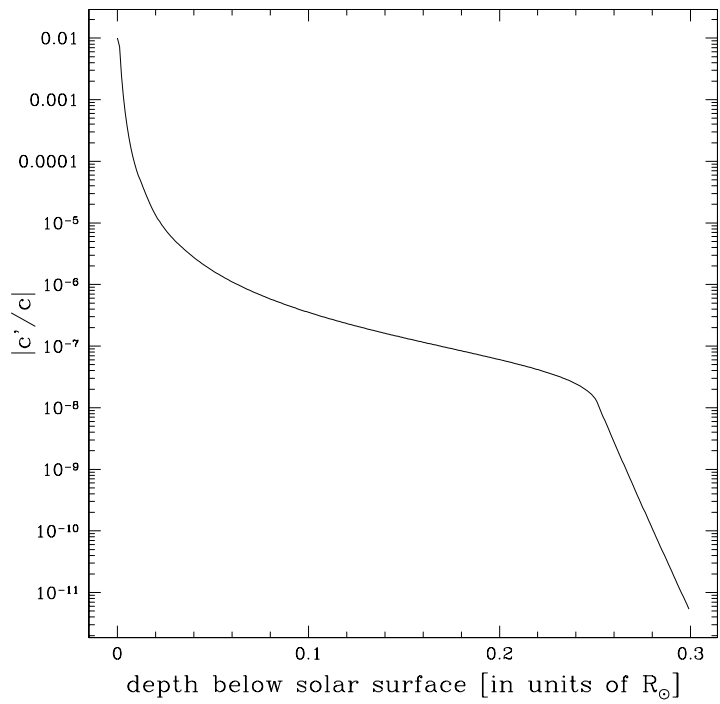
\begin{figure}
\caption{Sound speed fluctuations in the solar envelope.}
\label{deltac}
\end{figure}

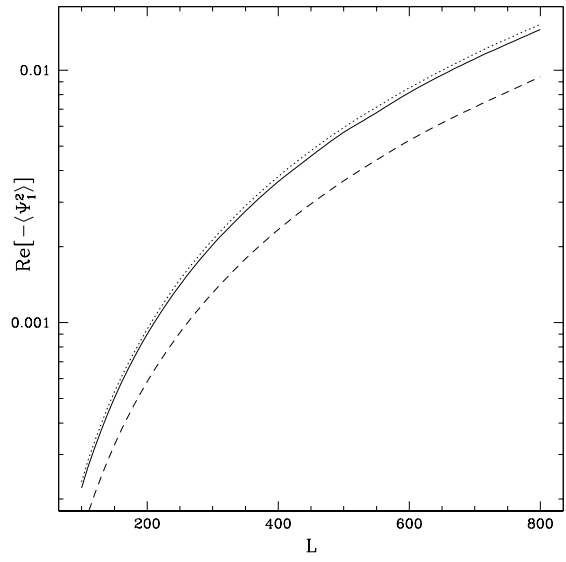
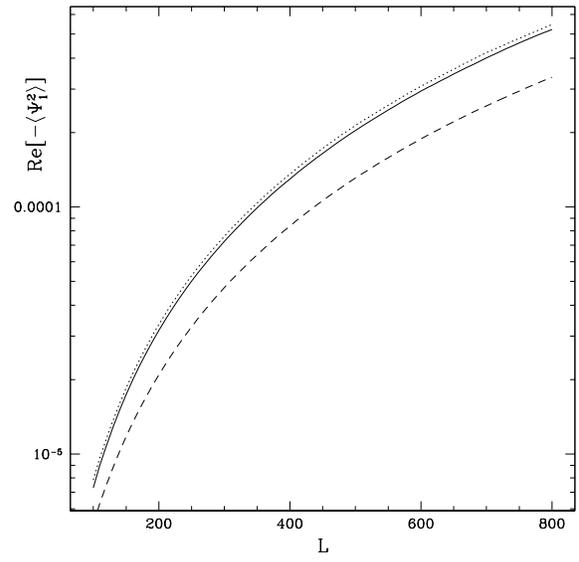
\begin{figure}
\caption{Real part of $-\langle\Psi_1^2\rangle$ as a function of $L$
for lower turning points of $a=0.01\,R_{\odot}$ (a) and
$a=0.001\,R_{\odot}$ (b) assuming different turbulent spectra (see
text).}
\label{amp1}
\end{figure}

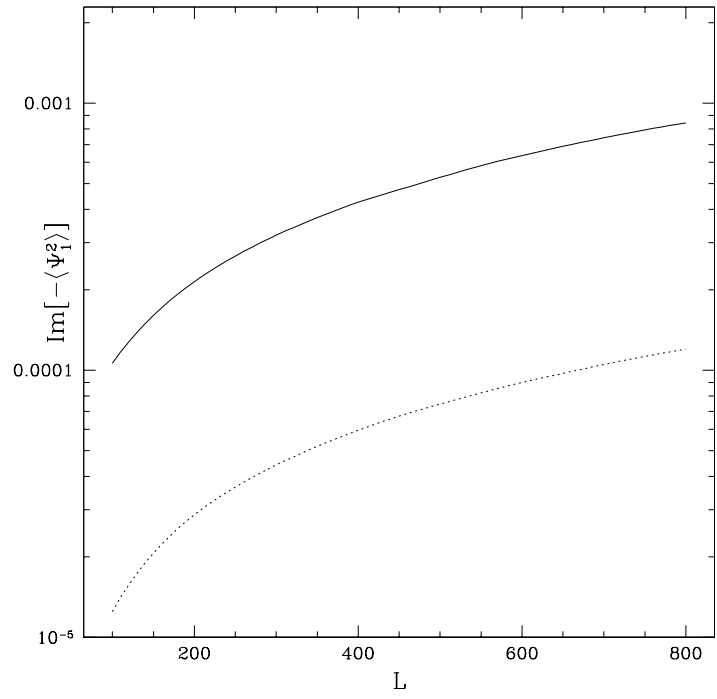
\begin{figure}
\caption{Imaginary part of $-\langle\Psi_1^2\rangle$ as a function of
$L$ for lower turning points of $a=0.01\,R_{\odot}$ (solid line) and
$a=0.001\,R_{\odot}$.}
\label{imaginary}
\end{figure}

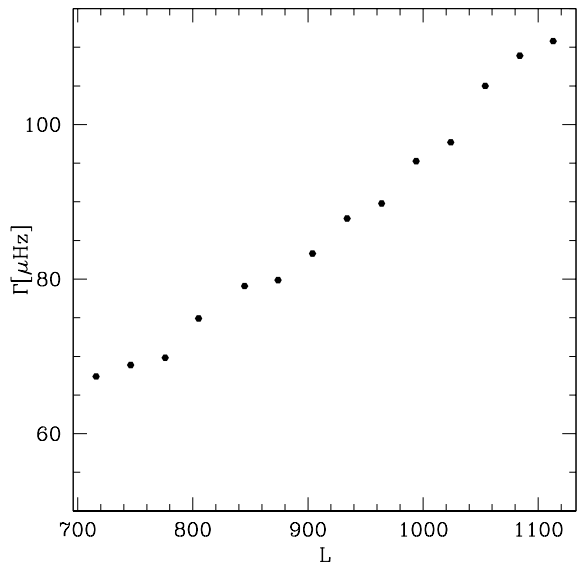
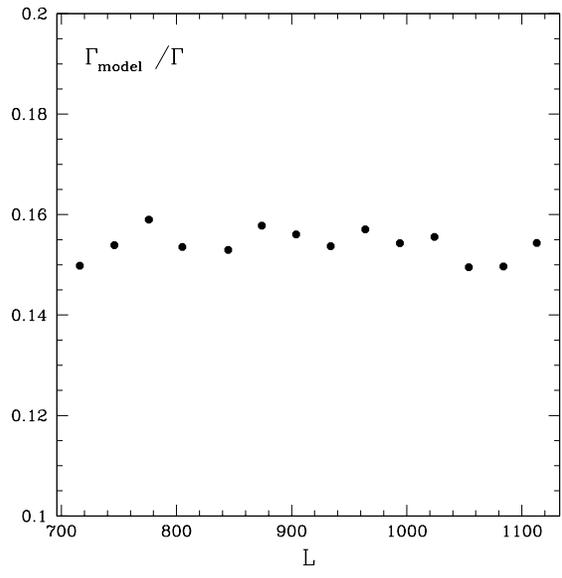
\begin{figure}
\caption{Panel (a) shows line widths of solar modes as measured by
Duvall, Kosovichev, \& Murawski (1998) using the MDI instrument. On
panel (b) are shown the corresponding (fractional) contributions to
the line width from sound speed scattering as predicted by my model.}
\label{linewidths}
\end{figure}

\label{lastpage}

\end{document}